\documentclass[10pt,article]{article}
\usepackage[top=0.85in,left=0.75in,footskip=0.75in,marginparwidth=2in]{geometry}
\usepackage{physics}
\usepackage{nameref,hyperref}
\usepackage{graphicx}
\usepackage{color}
\usepackage{abstract,lipsum}
\usepackage{cite}

\usepackage[aboveskip=1pt,labelfont=bf,labelsep=period,singlelinecheck=off]{caption}

\makeatletter
\renewcommand{\@biblabel}[1]{\quad#1.}
\makeatother

\usepackage{lastpage,fancyhdr,graphicx}
\usepackage{epstopdf}

\usepackage{wrapfig}
\usepackage[pscoord]{eso-pic}
\usepackage[fulladjust]{marginnote}
\reversemarginpar

\begin{document}
\vspace*{0.35in}

\begin{flushleft}
{\LARGE
\textbf\newline{ Absorption-free superluminal light propagation in a Landau-quantized graphene}
}
\newline
\\
Seyedeh Hamideh Kazemi\textsuperscript{1},
Mohammad Ali Maleki\textsuperscript{1},
Mohammad Mahmoudi\textsuperscript{1,*}
\\
\bigskip
\textsf{1} Department of Physics, University of Zanjan, University Blvd., 45371-38791, Zanjan, Iran
\\
* mahmoudi@znu.ac.ir

\end{flushleft}
\begin{abstract}
In recent years, control of group velocity of light has attracted enormous interest. One of the main challenges is to realize an absorption-free fast or slow light propagation. Here, we study dispersion and absorption properties of a weak probe field in a Landau-quantized graphene and report a gain-assisted superluminal light propagation. Moreover, an attempt is made to develop an analytical expression and necessary parameters for switching the group velocity of the probe field from subluminal to superluminal. It's worth mentioning that large dephasing rate in graphene offers feasibility of superluminal propagation of ultrashort light pulses. Additionally, dynamical behavior of dispersion and absorption of a weak probe field in a closed-type graphene system is investigated, and it is found that the absorption and dispersion can be dramatically affected by both the relative phase of applied fields and the Rabi frequencies in such a way that a large transient gain can be achieved and a transient absorption can be completely eliminated. 
\end{abstract}
\vspace*{0.2in}

\section{\label{sec:level1}Introduction}

Graphene, as the thinnest material known in the universe \cite{geim}, consists of carbon atoms in a two-dimensional (2D) hexagonal lattice with unusual Dirac-like electronic excitations. It holds many records related to mechanical, thermal, electrical and optical properties \cite{nobel,jens}. Besides, its band-gap structure can be tuned by voltage or chemical doping, through which conductivity and transmission are changed. Subsequently, this feature can endow graphene with a capability of operation in both terahertz and optical frequency ranges. In addition to the interest in fundamental research of optoelectronic and condensed matter physics \cite{geim,Katsnelson}, graphene is gaining attention owing to its various technological applications \cite{wang,xu1,bao,yan,Mueller}. Moreover, extant literature has reported investigations concerning the optical properties of graphene \cite{1,2,3,ding4,hamed,cox}. Not only do the investigations provide insights into the underlying nature of graphene's excited states, but they also open up interesting perspectives for emerging photonic and optoelectronic applications. Research has shown that magneto-optical properties of the graphene and thin graphite layer lead to multiple absorption peaks and unique selection rules for the allowed transitions \cite{sado,abergel}. Furthermore, non-equidistant Landau-levels (LLs) and the selection rules make graphene an excellent candidate for LL laser \cite{Aoki,Malic}. Moreover, its optical nonlinearity features have been exploited in multi-wave mixing \cite{1a,2a}, entangled photons \cite{yao2} and third harmonic generation \cite{chengnew}. These achievements demonstrate the feasibility of graphene for applications such as chip-scale high-speed optical communications, all-optical signal processing, photonics and optoelectronics. 

In the past three decades, controlling group velocity of light has attracted
a lot of interest owing to its potential applications, such as tunable optical buffers, optical memory and enhancing the nonlinear effect. Thus, numerous experimental and theoretical works have been devoted to control it in materials such as atomic medium \cite{boller,chiao2,kash,wang0,arbiv,menon,sahrai,evers,Davuluri}, alexandrite crystal \cite{boyd3}, optomechanical system \cite{safavi} and superconducting phase quantum circuit system \cite{sabegh}. Slow light can be used in telecommunication applications such as controllable optical delay lines, optical buffers \cite{2m}, true time delay methods for synthetic aperture radar, development of spectrometers with enhanced spectral resolution \cite{3m} and optical memories \cite{4m}. On the other hand, the question of wave velocity has been studied since the advent of Einstein's special theory of relativity \cite{velocity2,velocity1}. The key issue is whether the speed of light in vacuum, $c$, is an upper limit to the group velocity. Theoretical works \cite{Garrett,Crisp,Varoquaux} showed that the group velocity is not limited and a great deal of experiments confirmed that it is possible for optical or electrical wave pulses to travel through absorbing, attenuating or gain materials with group velocities greater than $c$ \cite{wong,chiao,krausz}. Another interesting scenario in light propagation concerns the situation where the group velocity of light can even become negative \cite{wang0,munday}. It is worth mentioning that the superluminal light propagation does not violate Einstein's theory of special relativity since the energy and information flow do not exceed $c$ \cite{chiao7,stenner}. By using the principle of such propagation, one can improve the speed of information transfer in telecommunication. An ideal condition for practical light propagation is a region in which the light pulse should not attenuate or amplify, primarily due to fact that pulse propagation does not possible in the presence of a large absorption. Besides, gain may add some noise to the system. Moreover, going towards superluminal propagation of shorter pulses is highly desirable; in this context, graphene potentially facilitates superluminal propagation of such pulses because of its large dephasing rate, about 30 $\mathrm{ps}^{-1}$, and high optical nonlinearity.

As mentioned above, considerable attention was paid to optical properties of graphene, however, superluminal light propagation in such system has received scant study to date. Despite the achievement of superluminal light propagation in graphene in only a few studies \cite{asadpour5,shiri}, it is accompanied by considerable absorption so that the pulse propagation does not possible, resulting in a drawback of practical applications. In this paper, we report a gain-assisted superluminal light propagation in a Landau-quantized graphene and show that the slope of dispersion can change from positive to negative just by adjusting the intensities of the coupling or controlling fields. Additionally, many works have focused on transient properties of probe field absorption, gain and enhancement of dispersion in both atomic and solid-state systems \cite{wei,xu,feng,asadpour}. Despite the importance of the transient behavior, there is a little study on this phenomenon in graphene \cite{heinz,knorr,hamedi7}. In fact, no study has been reported to date on transient optical properties of a closed-type graphene system. In present paper, we turn our attention to the role of the relative phase of applied fields on the transient optical properties of a graphene monolayer system. Motivated by a recent study on phase sensitivity of optical bistability and multistability in graphene \cite{dzhang}, we investigate the transient optical properties of the Landau-quantized graphene monolayer system interacting with three laser fields. The effects of both Rabi frequencies and relative phase of applied fields on the probe field absorption and dispersion are investigated. It is shown that the absorption and dispersion can be dramatically affected by the relative phase and the Rabi frequencies so that the transient absorption can be completely eliminated and a large transient gain can be achieved just by choosing the proper relative phase.

\section{\label{sec:1}Methods}
\begin{figure*}
\centering
\includegraphics[width=13.3cm]{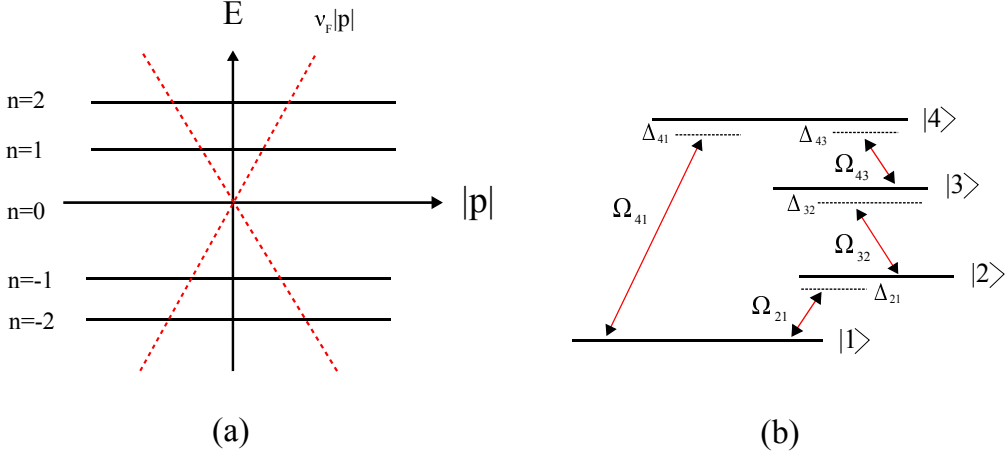}
\caption{ Graphene's energetically lowest LLs near Dirac point superimposed on electronic energy dispersion without applying a magnetic field $\pm \nu_{F} \vert p \vert$ (a). Energy levels and optical transitions in a Landau-quantized graphene interacting with three coherent fields. The states $\vert 1 \rangle$, $\vert 2 \rangle$, $\vert 3 \rangle$ and $\vert 4 \rangle$ correspond to the LLs with energy quantum numbers $n=-2,-1,0$ and $1$, respectively. Noting that fields are perpendicularly incident on the single-layer graphene where is treated as a perfect 2D crystal structure in the x-y plane (b).}
\label{figure 1}
\end{figure*}

Graphene, a one-atom thick allotrope of carbon, has a honeycomb hexagonal lattice structure \cite{nobel,1}. Unlike a conventional 2D electron/hole system with a parabolic dispersion $\varepsilon_{k} =k^2/(2m)$, graphene has a linear dispersion relation in the nearest-neighbor approximation and close to the Dirac points. In the presence of a perpendicular strong magnetic field ($B$), the LL energies are given by $\varepsilon_{n}=sgn(n) \hbar \omega_{c} \sqrt{\vert n \vert}$, where the integer $n$ is energy quantum number and $\omega_c=\sqrt{2} \nu_{F}/l_{B}$. Fermi velocity ($\nu_F$) is approximately $10^6$ m/s and $ l_{B}=\sqrt{\hbar/(eB)}$ means the magnetic length \cite{kim,ando}. These energy levels can be expressed as $\lambda_{c}[\mathrm{\mu m}]=34(B[$Tesla$])^{-1/2}$, in the wavelength scale and the energy, $\hbar\omega_c$, for $B\sim 1 \mathrm{T}$ is in the range of $CO_2$ laser. Note that transition between the LLs obeys the selection rule of a graphene monolayer system: $\bigtriangleup \vert n \vert = \pm 1$ \cite{abergel}. It may be noted that these unique selection rules enable transition with change in $n$ greater than $1$, as opposed to selection rules for electron: $\triangle n= \pm 1$. For instance, transition from $n=-1$ to $n=2$ is allowed leading to an efficient resonant nonlinear mixing. The external magnetic field condenses the original continuous states in the Dirac cone into discrete LLs which are proportional to $\sqrt{B}$, See Fig.~\ref{figure 1}(a). Note that LLs for graphene are unequally spaced, unlike levels in a conventional electron-hole system, $E_{n}=(n+1/2) \hbar e B / m$.

We consider a graphene monolayer system with four energy levels in presence of the strong magnetic field, as illustrated in Fig.~\ref{figure 1}(b). It is assumed that inter-Landau level transition $\vert 4 \rangle \leftrightarrow \vert 1 \rangle $ is driven by a right-handed circularly polarization field with the amplitude $E_{1}$ and the carrier frequency $\omega_{1}$, while another field with the amplitude $E_{2}$ and the carrier frequency $\omega_{2}$ interacts with transition $\vert 1 \rangle \leftrightarrow \vert 2 \rangle $. Such system has been already used for investigating the optical bistability behavior \cite{dzhang}. Third field can be written as $\vec{E_{3}}=(E_{3}^{+} \sigma ^{+} +E_{3}^{-} \sigma ^{-})e^{-i(\omega_{3}t-\vec{k}_{3}.\vec{r})} +c.c$, where $\vec{k}_{3}$, $ \sigma ^{-}$ and $ \sigma ^{+}$ are the wave vector, the unit vectors of the left- and right-hand circular polarization, respectively. Linearly polarized field with carrier frequency $\omega_{3}$ drives the transition $\vert 3 \rangle \leftrightarrow \vert 2 \rangle$ via $\sigma ^{+}$ component and the transition $\vert 4 \rangle \leftrightarrow \vert 3 \rangle $ is driven via $\sigma ^{-}$ component. Rabi frequencies of the corresponding fields are denoted by $\Omega_{41}=(E_{1} \vec{\mu_{41}}.\vec{e_{1}})/(2\hbar)$, $\Omega_{21}=(E_{2} \vec{\mu_{21}}.\vec{e_{2}})/(2\hbar)$, $\Omega_{32}=(E_{3}^{+} \vec{\mu_{32}}.\vec{e_{3}})/(2\hbar)$ and $\Omega_{43}=(E_{3}^{-} \vec{\mu_{43}}.\vec{e_{3}})/(2\hbar)$, with $\vec{\mu}_{ij}=e.\langle i\vert \vec{r} \vert j \rangle ,(i=2,3,4$ and $j=1,2,3)$ being as the electric dipole moment of the relevant transition. Noting that dipole moment of the transition between the LLs in
graphene has a magnitude of the order of $\vert \mu_{ij} \vert \sim e \hbar \nu_{F}/(\varepsilon_{j}-\varepsilon_{i}) \propto 1/\sqrt{B}$ \cite{abad}. For transitions near Dirac point ($\varepsilon_{j}-\varepsilon_{i} \sim \hbar \omega$), this is a large value. For example, the dipole moment at a magnetic field of $ 1 T$ falls into the mid-far infrared range: $ \vert \mu_{ij} \vert /e \sim 18\, \mathrm{nm}$.

In the absence of an optical field, the effective-mass Hamiltonian for a graphene monolayer in a magnetic field and in the nearest-neighbour tight-binding model can be written as a $4 \times 4$ matrix \cite{ando}:
 \begin{equation}
    \hat{H_{0}}= \nu_{F} \begin{bmatrix}
0 & \hat{\pi_{x}}-i \hat{\pi_{y}} & 0 & 0 \\
\hat{\pi_{x}}+i \hat{\pi_{y}} & 0 & 0 & 0 \\
0 & 0 & 0 & \hat{\pi_{x}}+i \hat{\pi_{y}} \\
0 & 0 & \hat{\pi_{x}}-i \hat{\pi_{y}} & 0
\end{bmatrix},
\end{equation}

where generalized momentum operator is denoted by $\hat{\pi}=\hat{P}+e \hat{A}/c$ with $\hat{P}$ and $e$ being as electron momentum operator and electron charge, respectively. Also, $\hat{A}$ is vector potential which is equal (0,\,$Bx$) for a constant magnetic field. In the presence of incident optical fields, we add vector potential of the optical field ($\vec{A}_{opt}=ic\vec{E}\omega$) to the vector potential of the magnetic field in the generalized momentum operator $\hat{\pi}$. Here, $\vec{E}$ is the sum of the incident optical fields: $\vec{E}=\vec{E}_{1}+\vec{E}_{2}+\vec{E}_{3}$. The resulting interaction Hamiltonian can be written as $\hat{H_{int}}=(\nu_{F} e/c) \vec{\sigma}. \vec{A}_{opt}$. Unlike the case of an electron with a parabolic dispersion relation, no higher order terms such as $\pi^2$ exist and the Hamiltonian is a  linear function of the vector potential $\vec{A}_{opt}$, even for strong optical fields. Moreover, the interaction Hamiltonian and its matrix elements are only decided by Pauli matrix vector $\vec{\sigma}=(\sigma_x,\sigma_y)$. 

It would be worth mentioning that graphene is not immune to disorder and thus to describe a more realistic situation, such sources need to be considered. It is found that more important sources of disorder give rise to a LL broadening, however, good agreement can be reached by assuming a constant and phenomenological rate, as can be learned from similar analysis \cite{jiang7,yao,liu6}. Furthermore, valley-degeneracy in the LL spectrum due to spin-orbit interaction and Zeeman splitting (spin-degeneracy)- both of which are small as compared to LL broadening in graphene- is superimposed by this energy broadening \cite{geo,wen1,wen2}. 
Now, a standard time-evolution equation for the density matrix of Dirac electrons in the graphene can be written as $d \hat{\rho}/dt =(1/i\hbar)[\hat{H}_{int},\hat{\rho}] +\hat{R}(\hat{\rho})$. The phenomenological decay rate $\hat{R}(\hat{\rho})= -1/2 \lbrace\hat{\Gamma},\hat{\rho}\rbrace=-1/2(\hat{\Gamma} \hat{\rho} + \hat{\rho} \hat{\Gamma})$  describes incoherent relaxation due to disorder, interaction of Dirac electrons with photons and carrier-carrier interactions. Moreover, the decay rate of the graphene is combined into the equation by a relaxation matrix, $\hat{\Gamma}$, where can be defined as  $\langle i \vert \hat{\Gamma} \vert j \rangle= \gamma_{i} \delta_{ij}$. Subsequently, a standard time-evolution equation for the density matrix of the system can be calculated as follows \cite{dzhang}:

\begin{subequations}
\begin{eqnarray}
\dot{\tilde{\rho}}_{44} &=& i \tilde{\Omega}_{41} \tilde{\rho}_{14}+ i \tilde{\Omega}_{43} \tilde{\rho}_{34}- i (\tilde{\Omega}_{41}) ^{*}\tilde{\rho}_{41} 
- i (\tilde{\Omega}_{43})^{*} \tilde{\rho}_{43}-\gamma_{4}\tilde{\rho}_{44}, \\
\dot{\tilde{\rho}}_{33} &=& i \tilde{\Omega}_{32} \tilde{\rho}_{23}+ i (\tilde{\Omega}_{43})^{*}\tilde{\rho}_{43}- i (\tilde{\Omega}_{32})^{*}\tilde{\rho}_{32} - i \tilde{\Omega}_{43} \tilde{\rho}_{34}-\gamma_{3}\tilde{\rho}_{33}, \\
\dot{\tilde{\rho}}_{22} &=& i \tilde{\Omega}_{21} \tilde{\rho}_{12}+ i (\tilde{\Omega}_{32})^{*}\tilde{\rho}_{32}- i (\tilde{\Omega}_{21})^{*}\tilde{\rho}_{21} -i \tilde{\Omega}_{32} \tilde{\rho}_{23}-\gamma_{2}\tilde{\rho}_{22}, \\
\dot{\tilde{\rho}}_{41} &=& i \tilde{\Omega}_{41} (\tilde{\rho}_{11}-\tilde{\rho}_{44})+ i \tilde{\Omega}_{43} \tilde{\rho}_{31}- i \tilde{\Omega}_{21} \tilde{\rho}_{42}
+(i\Delta_{41}- \dfrac{\gamma_{4}}{2})\tilde{\rho}_{41}, \\
\dot{\tilde{\rho}}_{42} &=& i \tilde{\Omega}_{41} \tilde{\rho}_{12}+ i \tilde{\Omega}_{43} \tilde{\rho}_{32}- i (\tilde{\Omega}_{21} )^{*}\tilde{\rho}_{41} -i \tilde{\Omega}_{32} \tilde{\rho}_{43}+( i\Delta_{43}+i\Delta_{32} -(\dfrac{\gamma_{4}+\gamma_{2}}{2}))\tilde{\rho}_{42}, \\
\dot{\tilde{\rho}}_{43} &=& i \tilde{\Omega}_{41} \tilde{\rho}_{13}+ i \tilde{\Omega}_{43} (\tilde{\rho}_{33}-\tilde{\rho}_{44}) 
- i (\tilde{\Omega}_{32})^{*}\tilde{\rho}_{42}+(i\Delta_{43}-(\dfrac{\gamma_{4}+\gamma_{3}}{2}))\tilde{\rho}_{43}, \\
\dot{\tilde{\rho}}_{31} &=& i (\tilde{\Omega}_{43})^{*}\tilde{\rho}_{41}+ i \tilde{\Omega}_{32} \tilde{\rho}_{21}- i \tilde{\Omega}_{21} \tilde{\rho}_{32}-i \tilde{\Omega}_{41} \tilde{\rho}_{34}+ (i\Delta_{41}+i\Delta_{32}-\dfrac{\gamma_{3}}{2})\tilde{\rho}_{31}, \\
\dot{\tilde{\rho}}_{32} &=& i \tilde{\Omega}_{32} (\tilde{\rho}_{22}-\tilde{\rho}_{33})+ i (\tilde{\Omega}_{43})^{*}\tilde{\rho}_{42} -i (\tilde{\Omega}_{21})^{*}\tilde{\rho}_{31}+(i\Delta_{32}-(\dfrac{\gamma_{2}+\gamma_{3}}{2}))\tilde{\rho}_{32}, \\
\dot{\tilde{\rho}}_{21} &=& i \tilde{\Omega}_{21} \tilde{\rho}_{11}+ i (\tilde{\Omega}_{32})^{*}\tilde{\rho}_{31}- i \tilde{\Omega}_{21} \tilde{\rho}_{22} 
-i \tilde{\Omega}_{41} \tilde{\rho}_{24}+( i \Delta_{21}-\dfrac{\gamma_{2}}{2})\tilde{\rho}_{21}.
\end{eqnarray}
\label{eq 4}
\end{subequations}

Here, overdots stand for the time derivations and the remaining equations follow from the constraints: $\tilde{\rho}_{lm}=\tilde{\rho}^{*}_{ml}$ with $l,m \in \lbrace 1,...,4 \rbrace$ and $\tilde{\rho}_{11}+\tilde{\rho}_{22}+\tilde{\rho}_{33}+\tilde{\rho}_{44}=1$. Also, the parameters $\Delta_{41}= \omega_{1}-(\varepsilon_{1}-\varepsilon_{-2})/\hbar$, $\Delta_{21}=\omega_{2} -(\varepsilon_{-1}-\varepsilon_{-2})/\hbar$, $\Delta_{32}=\omega_{3}-(\varepsilon_{0}-\varepsilon_{-1})/\hbar$ and $\Delta_{43}=\omega_{3}-(\varepsilon_{1}-\varepsilon_{0})/\hbar$ represent the corresponding detunings. Moreover, $\gamma_{i}$ corresponds to the decay rate of the state $\vert i \rangle$. 

\section{\label{sec:2}Results}
\subsection{Transient optical properties of a closed-type graphene monolayer system}
In this section, we assume that transitions $\vert 4 \rangle \leftrightarrow \vert 1 \rangle $ and $\vert 2 \rangle \leftrightarrow \vert 1 \rangle $ are driven by probe and coupling fields: $\tilde{\Omega}_{41}=\tilde{\Omega}_{p}, \tilde{\Omega}_{21}=\tilde{\Omega}_{c}$ with the corresponding detunings $\Delta_{41}=\Delta_{p}, \Delta_{21}=\Delta_{c}$. Further, transitions $\vert 4 \rangle \leftrightarrow \vert 3 \rangle $ and $\vert 3 \rangle \leftrightarrow \vert 2 \rangle $ are driven simultaneously by a control field: $\tilde{\Omega}_{43}=\tilde{\Omega}_{2}^{-}, \tilde{\Omega}_{32}=\tilde{\Omega}_{2}^{+}$ with $\Delta_{43}=\Delta_{32}=\Delta_{3}$. Then, we proceed to rewrite density matrix equations for the case of a closed-loop configuration, in which the system becomes quite sensitive to the relative phase of applied fields. Taking $\varphi_{p}$, $\varphi_{c}$ and $\varphi_{3}$ as phases of probe, coupling and control fields,  treating the Rabi frequencies as complex-valued parameters: $\tilde{\Omega}_{p}=\Omega_{p} e^{-i\varphi_{p}}$, $\tilde{\Omega}_{c}=\Omega_{c} e^{-i\varphi_{c}}$, $\tilde{\Omega}^{+}_{2}=\Omega^{+}_{2} e^{-i\varphi_{3}}$ and $\tilde{\Omega}^{-}_{2}=\Omega^{-}_{2} e^{-i\varphi_{3}}$, and therefore redefining the density matrix elements: $\tilde{\rho}_{41}=\rho_{41} e^{-i\varphi_{p}}$, $\tilde{\rho}_{42}=\rho_{42} e^{-i\varphi_{3}}$, $\tilde{\rho}_{43}=\rho_{43} e^{-i\varphi_{3}}$, $\tilde{\rho}_{31}=\rho_{31} e^{-i(\varphi_{p}-\varphi_{3})}$, $\tilde{\rho}_{32}=\rho_{32} e^{-i\varphi_{3}}$ and $\tilde{\rho}_{21}=\rho_{21} e^{-i(\varphi_{p}-2 \varphi_{3})}$, we can obtain equations for the density matrix elements:

\begin{subequations}
\begin{alignat}{9}
\dot{\rho}_{44} &= i \Omega_{p} \rho_{14}+ i \Omega_{2}^{-}\rho_{34}- i \Omega_{p} ^{*}\rho_{41}
-i (\Omega_{2} ^{-})^{*}\rho_{43}- \gamma_{4}\rho_{44}, \\
\dot{\rho}_{33} &= i \Omega_{2}^{+} \rho_{23}+ i (\Omega_{2}^{-} )^{*}\rho_{43}- i (\Omega_{2}^{+} )^{*}\rho_{32}
- i \Omega_{2}^{-} \rho_{34}-\gamma_{3}\rho_{33}, \\
\dot{\rho}_{22} &= i \Omega_{c} e^{i\varphi} \rho_{12} + i (\Omega_{2}^{+} )^{*}\rho_{32}
-i (\Omega_{c} )^{*}e^{-i\varphi}\rho_{21}-i \Omega_{2}^{+} \rho_{23}-\gamma_{2}\rho_{22}, \\
\dot{\rho}_{41} &= i \Omega_{p} (\rho_{11}-\rho_{44})+ i \Omega_{2}^{-} \rho_{31}- i \Omega_{c} e^{i\varphi} \rho_{42}
+(i\Delta_{p}-\dfrac{\gamma_{4}}{2} )\rho_{22}, \\
\dot{\rho}_{42} &= i \Omega_{p} \rho_{12}+ i \Omega_{2}^{-}\rho_{32}- i (\Omega_{c} )^{*} e^{-i\varphi}\rho_{41}
-i \Omega_{2}^{+}\rho_{43}+(2 i\Delta_{3}-(\dfrac{\gamma_{4}+\gamma_{2}}{2}))\rho_{42}, \\
\dot{\rho}_{43} &= i \Omega_{p} \rho_{13}+ i \Omega_{2}^{-} (\rho_{33}-\rho_{44})- i (\Omega_{2}^{+} )^{*}\rho_{42}
+(i\Delta_{3}-(\dfrac{\gamma_{4}+\gamma_{3}}{2}))\rho_{43}, \\
\dot{\rho}_{31} &= i (\Omega_{2} ^{-})^{*}\rho_{41}+ i \Omega_{2}^{+} \rho_{21}- i \Omega_{c} e^{i\varphi}\rho_{32}
-i \Omega_{p} \rho_{34}+(i\Delta_{p}+ i\Delta_{3}-\dfrac{\gamma_{3}}{2} )\rho_{31}, \\
\dot{\rho}_{32} &= i \Omega_{2} ^{+} (\rho_{22}-\rho_{33})- i (\Omega_{c})^{*}e^{-i\varphi}\rho_{31}
+ i (\Omega_{2}^{-} )^{*}\rho_{42}+(i\Delta_{3}-(\dfrac{\gamma_{2}+\gamma_{3}}{2}))\rho_{32}, \\
\dot{\rho}_{21} &= i\Omega_{c} e^{i\varphi}\rho_{11}+ i (\Omega_{2}^{+})^{*}\rho_{31}- i \Omega_{c} e^{i\varphi} \rho_{22}
-i \Omega_{p} \rho_{24}+(i\Delta_{c}-\dfrac{\gamma_{2}}{2})\rho_{21},
\end{alignat}
\label{eq 5}
\end{subequations}

where the parameter $\varphi=\varphi_{p}-2\varphi_{3}-\varphi_{c}$ is the relative phase of applied fields. It is also assumed that conditions $\omega_{p}=\omega_{c}+2 \omega_{3}$ and $\Delta_{p}=\Delta_{c}+2 \Delta_{3}$ are fulfilled.

Before discussing the transient behavior of absorption and dispersion, we briefly review a recent research on the transient behavior of graphene. Hamedi and Sahrai have studied the evolutional absorption behavior of a Landau-quantized graphene structure and have investigated the impact of intensity and frequency detuning of driving fields on temporal evolution of probe absorption \cite{hamedi7}. Here, we will show the possibility of controlling such behavior by the relative phase of applied fields. Note that phase controlling of the phenomena is easier than that via the intensity or the frequency. The main advantage of our proposed scheme, therefore, is its simple implementation. From the experimental point of view, the relative phase could be easily changed by electro-optical devices. What is more, our model shows a wide range of tunability so that a large transient gain and steady-state absorption can be achieved, compared to their suggested scheme.
\begin{figure}
\centering
\includegraphics[width=13.3cm]{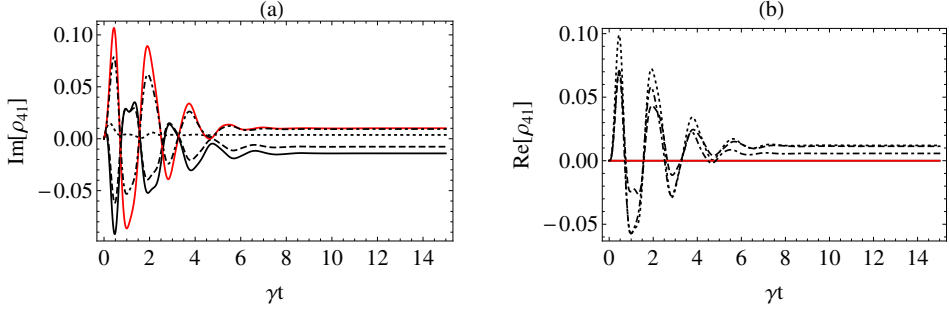}
\caption{ Temporal evolution of response of graphene to the probe field for same-order driving fields. The transient behavior of the absorption-dispersion is plotted for different relative phases: $\varphi=0$ (solid line), $\varphi=\pi/4$ (dashed line), $\varphi=\pi/2$ (dotted line), $\varphi=3\pi/4$ (dot-dashed line) and $\varphi=\pi$ (red line). Imaginary (a) and real (b) parts of $\rho_{41}$ are plotted for $\Delta_{3}=\Delta_{p}=0$, $\Omega_{p}=0.1 \gamma$ and $\Omega_{2}^{+}=\Omega_{2}^{-}=\Omega_{c}=3 \gamma $.}
\label{figure 2}
\end{figure}

Here, our main observable is the response of the medium to the probe field. Gain or absorption coefficient of the probe laser field coupled to the transition $\vert 4 \rangle \leftrightarrow \vert 1 \rangle $ is characterized with Im$[\rho_{41}]$, while the dispersion is proportional to Re$[\rho_{41}]$. In our numerical calculation, we take the transition frequency $\omega_{41} \sim 10^{14} \mathrm{s}^{-1}$ at the magnetic field of $1\sim 3 \mathrm{T}$ and assume $\gamma_{4}=\gamma_{3}=\gamma_{2}=\gamma=3 \times 10^{13} \mathrm{s}^{-1}$. It should be noted that the value assigned to $\gamma_{k}$ is a rather conservative choice considering the latest experimental and numerical works \cite{yao2,jiang7}. It is also assumed that the system is initially at the ground state. In our notation, Im$[\rho_{41}] < 0$  means that the system exhibits gain, while the probe field is attenuated when we have Im$[\rho_{41}] > 0$.

Now we turn on the discussing the transient behavior of the absorption and dispersion, based on the solution of the density matrix equations of the motion (Eqs.~(\ref{eq 5})), which are shown in Fig.~\ref{figure 2} for $\Delta_{3}=\Delta_{p}=0$, $\Omega_{p}=0.1 \gamma$, $\Omega_{2}^{+}=\Omega_{2}^{-}=\Omega_{c}=3 \gamma $ and for different relative phases.

For a weak probe field and the same-order driving fields, the probe absorption is phase dependent. At time $t=0$, the imaginary part of $\rho_{41}$ is zero. Once time increase, the absorption oscillates with a fast-damped amplitude, however, it will finally reach a steady-state. For $\varphi=0$ and $\varphi=\pi/4$, the steady-state values are negative, while for other relative phases they are positive. The dispersion properties of the weak probe field is also phase dependent, as can be seen in Fig.~\ref{figure 2}(b). For $\varphi=0$ and $\varphi=\pi$, dispersion response is zero, while for $\varphi=\pi/4$, $\varphi=\pi/2$ and $\varphi=3\pi/4$, we can see an oscillation signal in dispersion with small positive steady-state values.
\begin{figure}[!h]
\centering
\includegraphics[width=13.3cm]{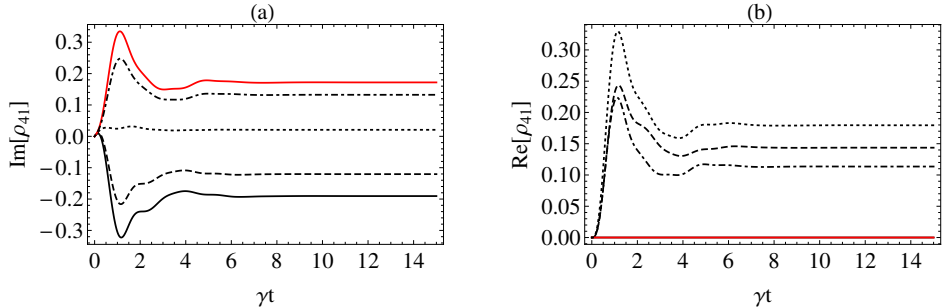}
\caption{Temporal evolution of response of graphene to the probe field: imaginary (a) and real (b) parts of $\rho_{41}$ are plotted for $\Omega_{2}^{+}=\Omega_{2}^{-}=3 \gamma $ and $\Omega_{c}=\gamma$. Other parameters are the same as in Fig.~\ref{figure 2}. }
\label{figure 3}
\end{figure} 

Fig.~\ref{figure 3} shows imaginary and real parts of $\rho_{41}$ for $\Omega_{2}^{+}=\Omega_{2}^{-}=3 \gamma $ and $\Omega_{c}=\gamma$. As can be seen, transient properties are greatly changed in such a way that they exhibit different features for different relative phases. In the presence of a strong control field, the results no longer exhibit periodic amplification and absorption. For $\varphi=0$ and $\varphi=\pi/4$, the transient absorption completely disappear, leaving a large transient gain. However, for other values of the relative phase, the absorption oscillates above the zero-absorption line and reaches a positive value. We can therefore achieve a larger gain just by decreasing the coupling field and choosing proper values of the relative phase.

\begin{figure}[t]
\centering
\includegraphics[width=13.3cm]{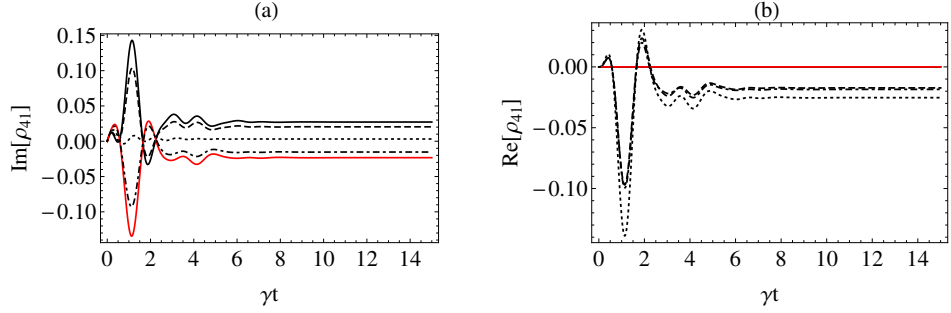}
\caption{Temporal evolution of response of graphene to the probe field: imaginary (a) and real (b) parts of $\rho_{41}$ are plotted for $\Omega_{2}^{+}=\Omega_{2}^{-}= \gamma $ and $\Omega_{c}=3 \gamma$. Other parameters are the same as in Fig.~\ref{figure 2}. }
\label{figure 4}
\end{figure}

In Fig.~\ref{figure 4}, we plot temporal evolution of response of graphene to a weak control field, $\Omega_{2}^{-}=\Omega_2^{+}= \gamma $. It can be seen that the transient properties is different from the previous one; transient absorption and dispersion can be found during the process. Also, the behavior of the imaginary and real parts of $\rho_{41}$ is similar to the situation of the same-order driving fields, except that the oscillatory frequency becomes smaller.

Before ending this section, we would like to make the following remark. As the graphene monolayer system suggested in this section is a closed-type one, a similar phase-sensitive behavior is also observed for all chosen parameters under the multi-photon resonance condition; indeed, the only constraint that must be satisfied by the parameters is that upon the detunings: $\Delta_{p}=\Delta_{c}+2 \Delta_{3}$. So, the result is not basically changed and the similar trend can be found with parameters different from those of Figs.~2-4.  

\subsection{Gain-assisted superluminal light propagation}

In this subsection, we assume that one of the fields is switched off ($\Omega_{21}=0$) and we treat $\Omega_{43}$, $\Omega_{41}$ and $\Omega_{32}$ as probe, coupling and controlling fields, respectively. The susceptibility of the weak probe field $\chi(\omega_{p})$ can be written as
\begin{equation}
\chi(\omega_{p})=\dfrac{N \mu_{34}^2}{\epsilon_{r} \hbar \Omega_{43}} \rho_{43},
\end{equation}
where $N$ and $\epsilon_{r}$ are sheet electron density of graphene and substrate dielectric constant, respectively.

We then introduce group index, $n_{g}=c/v_{g}$, where group velocity of the probe field is given by \cite{ficek}
\begin{equation}
v_{g}=\dfrac{c}{1+\dfrac{1}{2} \left[ Re[\chi(\omega_{p})]+ \omega_{p} \dfrac{\partial Re[\chi(\omega_{p})] }{\partial \omega_{p} } \right]}.
\label{eq 8}
\end{equation} 
As can be seen, for a negligible absorption the group velocity can be significantly reduced via a steep positive dispersion, while the strong negative dispersion can increase the group velocity to establish even a negative group velocity. For the present system, the electron density is assumed to be $N\simeq 55\times 10^{13} \mathrm{cm}^{-2}$, the dielectric constant turns out to be $\epsilon_{r}=4.5$ and the dipole moment between the transition $\vert 4 \rangle \leftrightarrow \vert 3 \rangle $ has a magnitude of the order of $1/ \sqrt{B}$.

Before presenting our results, we drive an analytical expression and the necessary parameters for switching the group velocity of the probe field from subluminal to superluminal. In the case of $\Delta_{41}=0$, $\Delta_{43}=\Delta_{32}=\Delta_{p}$ and $\Omega_{41} < \Omega_{32}$, the expression for the steady-state coherence, $\rho_{43}$, yields:

\begin{equation}
\rho_{43}= - \dfrac{ 8 i  (\gamma-2 i \Delta_{p} ) \, \Omega_{43} \vert \Omega_{41}\vert ^2  a_{0}}{(\gamma + 8 \vert \Omega_{41}\vert ^2 ) \left[  d_{0}- d_{1}  \Delta_{p} - d_{2} \, \Delta_{p}^2 -36 \,i \, \Delta_{p}^3 + 16 \, \Delta_{p}^4  \right]    },
\label{eq 9} 
\end{equation}
where we define $a_{0}=\gamma^{2}- 5 i  \gamma \Delta_{p} - 4 \Delta_{p}^2 + 2 \vert \Omega_{32}\vert ^2$,  $d_{0}=\gamma^{4}  + \gamma^{2} \left( 5  \vert \Omega_{32} \vert^{2} + 2 \vert \Omega_{41} \vert^{2}  \right) + 4 \vert \Omega_{32}\vert^{2}  $, $d_{1}= 9 i \gamma^{3}  + 6 i \gamma \left(  3 \vert \Omega_{32}\vert ^2 + 2 \vert \Omega_{41}\vert ^2 \right) $ and $d_{2}= 28 \gamma^2 + 16  \left(  \vert \Omega_{32}\vert ^2 + \vert \Omega_{41}\vert ^2 \right) $.
 
\begin{figure}[!h]
\centering
\includegraphics[width=8cm]{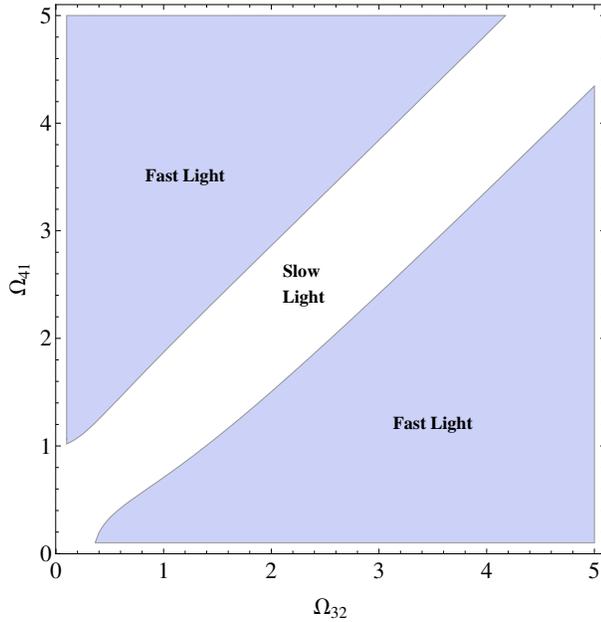}
\caption{ Sub- and superluminal regions are plotted via the Rabi frequency of controlling and coupling fields, around zero detuning.}
\label{figure 5}
\end{figure}

 Based on Eq.~(\ref{eq 9}), critical value of the controlling field at which the slope of dispersion changes from positive to negative is $ \Omega_{cr}= (1-\Omega_{41}^2)^{1/6} / \sqrt{2}$. For $\Omega_{32} < \Omega_{cr}$, the slope of dispersion around zero probe detuning is positive, while for $\Omega_{32} > \Omega_{cr}$ it becomes negative. In Fig.~\ref{figure 5}, we display subluminal and superluminal regions via the Rabi frequencies of the controlling and coupling fields for $\Omega_{43}=0.01 \gamma$. The superluminal regions are shown by dark color.

\begin{figure*}
\centering
\includegraphics[width=13.3cm]{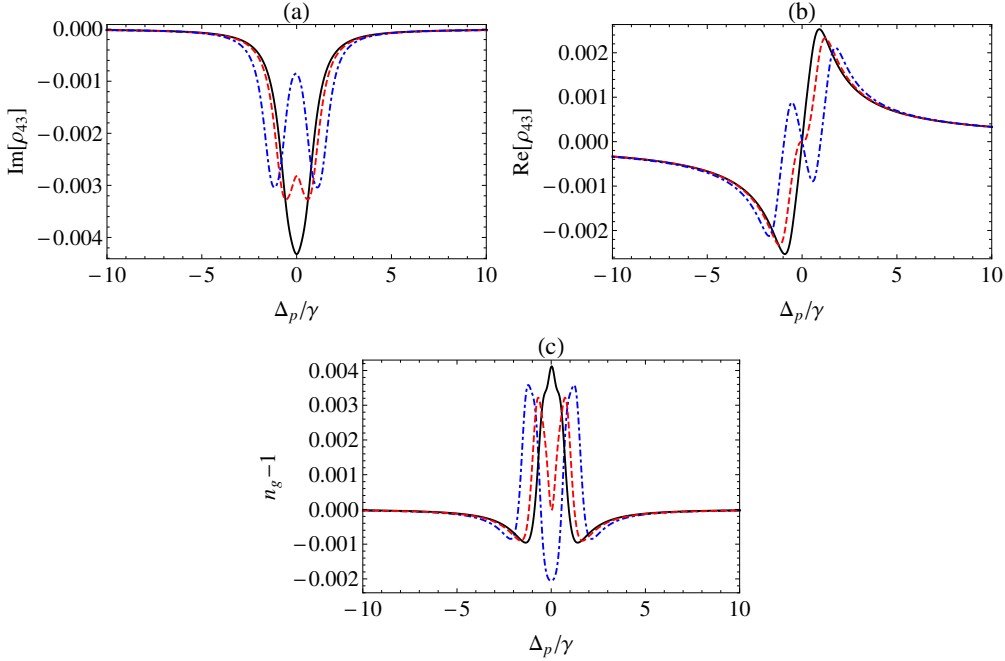}
\caption{Imaginary (a), real (b) parts of $\rho_{43}$ and group index (c) as a function of the probe detuning. The parameters are: $\Omega_{43}=0.01 \gamma$, $\Omega_{41}=0.5 \gamma$, $\Omega_{32}=0.2 \gamma$ (solid line), $\Omega_{32}=0.7 \gamma$ (dashed line) and $\Omega_{32}=1.5 \gamma$ (dot-dashed line). Other parameters are the same as in Fig.~\ref{figure 2}.}
\label{figure 6}
\end{figure*}

As the linear susceptibility of the weak probe field is determined by the coherence $\rho_{43}$, we therefore proceed with solving Eqs.~(\ref{eq 4}) in the steady-state situation, which can be obtained from those equations for vanishing time derivatives. In Figs.~\ref{figure 6}(a) and (b), we show imaginary and real parts of $\rho_{43}$ versus the probe detuning ($\Delta_{p}$) for various intensities of the controlling field ($\Omega_{32}$). In this figure, $\Omega_{21}$ is switched off, $\Omega_{43}=0.01 \gamma$, $\Omega_{41}=0.5 \gamma$ and other parameters are the same as in Fig.~2. As Fig.~\ref{figure 6}(a) shows, for all values of the controlling field, the system shows a gain structure. Furthermore, the initial gain separates into two dips by increasing the intensity of the controlling field. 

In Fig.~\ref{figure 6}(b), we see how the superluminal light propagation can be established for a specific value of the controlling field. It can be seen that when the intensity of the controlling field is sufficiently small, the slope of dispersion is positive corresponding to the subluminal light propagation. By increasing the intensity of the controlling field, when it reaches a critical value $\Omega_{cr}$, the nature of the curve changes so that it has a zero slope at $\Delta_{p}=0$. However, for $\Omega > \Omega_{cr}$ the slope of dispersion becomes negative. Note that the accuracy
of the analytical solution has been checked by comparing the analytical expression with numerical results, revealing a good agreement in terms of the critical value. For instance, the value is $\Omega_{cr}=0.7$ for the parameters of Fig.~\ref{figure 6}, which is in satisfactory agreement with the analytical prediction (about $0.67$).

In Fig.~\ref{figure 6}(c), we display the group index, $c/v_{g}-1$, versus the probe field detuning. It can be realized that for $\Omega_{32}=0.2 \gamma$, the group index around the zero detuning is positive, corresponding to the subluminal light propagation. For $\Omega_{32}=0.7 \gamma$, the group index around $\Delta_{p}$ is zero and the group velocity is equal to the speed of light in vacuum. By increasing the intensity of controlling field, the group index becomes negative, that is, the superluminal light propagation can be achieved. It should be pointed out that the gain-assisted superluminal light propagation in our scheme is accompanied by population inversion, unlike the case of electromagnetically induced transparency (EIT), where a gain doublet induced by quantum interference is established without population inversion. Note that the origin of the gain-doublet in our case is level splitting due to the dynamical Stark effect. 

 It is also worth comparing our scheme with a previous one involving optical properties of graphene monolayer nanostructure \cite{asadpour6}. The obvious difference is that the superluminal light propagation in our system is accompanied with a doublet gain to ensure that the light pulse does not attenuate as it passes through the medium, whereas in the proposal by Jamshidnejad \textit{et al.}, the superluminal light propagation is accompanied by amplification. From the viewpoint of practical applications, the superluminal light propagation produced in their study is not suitable. 
\begin{figure}
\centering
\includegraphics[width=14.3cm]{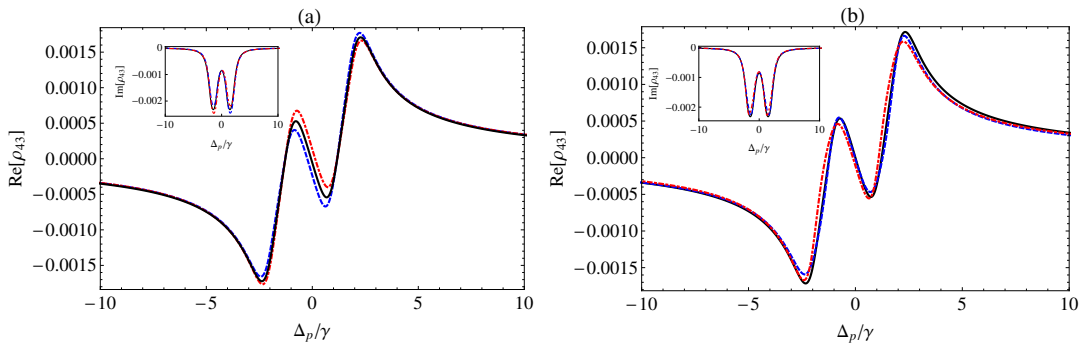}
\caption{Real part of $\rho_{43}$ as a function of the probe detuning for different detunings of the coupling field: $\Delta_{32}=0$ (solid line), $\Delta_{32}=0.2 \gamma$ (dashed line) and $\Delta_{32}=-0.2 \gamma$ (dot-dashed line) (a) and for different detunings of the controlling fields: $\Delta_{41}=0$ (solid line), $\Delta_{41}=0.2 \gamma$  (dashed line) and  $\Delta_{41}=-0.2 \gamma$ (dot-dashed line) (b). Imaginary part of  $\rho_{43}$ as a function of the probe detuning is shown in inset. Other parameters are the same as in Fig.~\ref{figure 6}.}
\label{figure 7}
\end{figure}

For practical applications, it is important to investigate the robustness of the scheme with respect to detunings and Rabi frequencies. The issue of dependence of subluminal and superluminal regions on controlling and coupling fields is addressed in Fig.~5. We also recalculate our results on light propagation for different detunings of controlling field ($\Delta_{32}$) and nonzero detunings of coupling field ($\Delta_{41}$) and find that the superluminal light propagation can be also achieved for those parameters in the weak-probe field regime. It is also imperative to point out that the constraint presented above is mainly due to the fact that a compact analytical solution can be obtained and similar behavior can be found with parameters different from those in Figs.~5 and 6. The only constraint that must be satisfied to produce the superluminal light propagation is that on the controlling and coupling fields presented in Fig.~5, which can readily be produced. 

In relation to Fig.~\ref{figure 6}, we find a further advantage of our suggested scheme; the insensitivity to fluctuation in detunings in such a way that both gain and slope of the dispersion at $\Delta_{p}$ remain almost unchanged. Fig.~\ref{figure 7} presents real part of $\rho_{43}$ as a function of the probe detuning for three different detunings:  $\Delta_{32}=0, \pm 0.2 \gamma$ (a) and $\Delta_{41}=0, \pm 0.2 \gamma$ (b). Insets in this figure depict the imaginary part of $\rho_{43}$. This figure clearly shows that the slope of dispersion remains almost unchanged for the above-mentioned variation in the detunings. It is also imperative to point out that the gain changes very slightly as the detunings fluctuate by $\pm 20$\% over the previous values. It is worth mentioning that the value assigned to the detuning (a few THz) are chosen according to a recent work on optical analogue of EIT in graphene \cite{gao}.

\section{Conclusion}
 In summary, we investigated the dispersion and absorption properties of a weak probe field in the Landau-quantized graphene monolayer system. We found that the slope of dispersion can be changed from positive to negative just by adjusting the intensity of coupling or controlling field and showed that this configuration allows for the gain-assisted subluminal and superluminal light propagation. The analytical expression and the necessary parameters for switching the group velocity  from subluminal to superluminal were also derived. It is worthy of mention that large dephasing rate in graphene suggests feasibility of superluminal propagation of ultrashort light pulses. In addition, the transient optical properties of a weak probe field in a closed-type graphene system was investigated. The effects of Rabi frequencies and relative phase of applied fields on the probe field absorption and dispersion were also investigated. It was shown that the dispersion and absorption were dramatically changed by the relative phase; in the case of a weak coupling field, a large transient gain can be achieved and the transient absorption can be completely eliminated just by choosing the proper relative phase. \newline

\bibliographystyle{}

\end{document}